\documentclass[twocolumn,showpacs,preprintnumbers,amsmath,amssymb]{revtex4}
\usepackage{dcolumn}
\usepackage{bm}
\usepackage{graphicx,subfigure,xcolor}
\usepackage[applemac]{inputenc}
\usepackage[T1]{fontenc}
\usepackage{amsmath,amssymb,amsfonts}
\usepackage{indentfirst}
\usepackage{graphics}
\usepackage{microtype}
\usepackage{caption}
\usepackage{multirow}
\usepackage{booktabs}
\usepackage{fancyhdr}
\usepackage{braket}
\usepackage{mathtools}
\usepackage{subfigure}
\usepackage{natbib}

\begin{document}

\title{State-Transfer Simulation in Integrated Waveguide Circuits}

\author{L. Latmiral$^1$}
\email{l.latmiral14@imperial.ac.uk}
\author{C. Di Franco$^1$}

\author{P. L. Mennea$^2$}

\author{M. S. Kim$^1$}

\affiliation{$^1$QOLS, Blackett Laboratory, Imperial College London, London SW7 2BW, United Kingdom\\$^2$Optoelectronics Research Centre, University of Southampton, Highfield SO17 1BJ, United Kingdom}


\begin{abstract}
Spin-chain models have been widely studied in terms of quantum information processes, for instance for the faithful transmission of quantum states. Here, we investigate the limitations of mapping this process to an equivalent one through a bosonic chain. In particular, we keep in mind experimental implementations, which the progress in integrated waveguide circuits could make possible in the very near future. We consider the feasibility of exploiting the higher dimensionality of the Hilbert space of the chain elements for the transmission of a larger amount of information, and the effects of unwanted excitations during the process. Finally, we exploit the information-flux method to provide bounds to the transfer fidelity.
\end{abstract}

\maketitle

\section{\label{sec:level1}Introduction}

Thanks to the development of modern technologies, quantum information and quantum communication are attracting increasing interest. In this context a leading idea is to exploit the interaction of elements in a many-body register. In particular, the {\it always-on} spin-chain model (a one-dimensional array of two-level quantum systems interacting through constant couplings) has been studied in detail, due to the promise of being a less challenging scenario in terms of experimental realizations. It has been proved in theory that this system can be exploited for many tasks, from the transmission of quantum states \cite{bose2003, bose2007} to the implementation of quantum computing \cite{benjamin2003}, among others. The influence that the dynamics of a particular element exerts on other parties during the time evolution of quantum many-body systems has been named \emph{information flux} \cite{Difranco2007, Difranco2008a}. Given a certain Hamiltonian and the distribution of coupling strengths, by operating on the initial state of the system one can control the information flux and thus the dynamics. Its investigation has also allowed the design of protocols where the initialization of the medium is not required \cite{Difranco2008, Difranco2009}. Unfortunately, many theoretical results obtained so far have not been demonstrated in experiments yet. This is due to the fact that, even if less challenging, the requirements for their implementations in the current setups are still demanding in terms of resources and control.

With the recent progress obtained in integrated circuits it has become feasible to write optical waveguides in glass where photons can be manipulated and interfere while preserving their coherence and polarization \cite{gattass2008flm, dellavalle2009, Sans2010, smith2009, lepert2011}. They have been successfully used for different applications, ranging from quantum walks \cite{Peru2010, Sans2012, owens2011} to boson sampling \cite{Broome2013, Spring2013, Till2012, Crespi2012}. These integrated waveguide circuits could also pave the way for experimentally demonstrating the protocols put forward for spin-chain models. Even if the photonic system evolves under the action of a different Hamiltonian, there are cases where a map to a chain of two-level systems is possible, and preliminary studies are currently in progress \cite{bellec2012, perez2013}. Providing an easily controllable setup as a proper benchmark for the theory will surely boost the investigation on spin-chain dynamics. In turn it would also be interesting, starting from the results already known, to generalize and modify or adapt the schemes to exploit the distinctive features of the bosons. As an indicative example, we will study in this paper the generalization of the state transfer to the case of a three-level quantum state, allowed by the larger Hilbert space of the single elements of our chain.

It is important not to forget the bosonic nature of the systems also when we are reproducing/simulating the results obtained for chains of spins. This clearly has an effect on the efficiency of the protocols. For instance, a non perfect initialization of the medium used for the state transfer could allow more excitations to be present during the time evolution. As we will show in this paper, the behavior will thus depend on the quantum nature of the elements of the chain.

After introducing in Sec.~\ref{sec2} the model considered in our investigation, we study in Sec.~\ref{sec3} the transmission of a general qutrit state in a bosonic chain; Sec. ~\ref{sec4} deals with the different results obtained using a bosonic chain instead of a standard spin chain, when unwanted excitations could be present in the medium; in Sec.~\ref{sec5} we exploit the information-flux approach to provide a bound for the correction on transfer fidelity; finally, we summarize the results in Sec.~\ref{sec6}.

\section{\label{sec2}The setup: a linear chain of coupled waveguides}
In this section we would like to briefly introduce the experimental system we will refer to in our study.
Henceforth, we are going to discuss the transmission of information through a linear chain of $N$ coupled sites, whose straightforward experimental implementation could be the evolution of photons in an array of $N$ evanescently coupled waveguides. The Hamiltonian that describes a chain of bosons with pairwise interactions reads
\begin{equation}
\mathcal{\hat{H}}=\sum_{i=1}^{N-1}k_i(\hat{a}_i\hat{a}^\dag_{i+1}+\hat{a}_{i+1}\hat{a}^\dag_i),
\label{boson_hamiltonian}
\end{equation}
where $\hat{a}_i$ and $\hat{a}^\dag_i$ are the annihilation and creation operators corresponding to channel $i$, and $k_i$ is the strength of the coupling between channels $i$ and $i+1$. We will mainly focus on two setups characterized by different coupling strengths $k_i$, experimentally controlled by the spacing between waveguides:
\begin{itemize}
\item the ideal case, in which couplings verify a perfect mirror-symmetry law: $k_i=J\cdot\sqrt{i\cdot(N-i)}$, where $J$ is a characteristic of the energy scale, thus depending on the physical implementation of the model;
\item an easier experimental configuration which corresponds to choosing all couplings $k_i=J$ for $i\in [2,N-2]$ and $k_1=k_{N-1}=K$.
\end{itemize}

While the former is clearly interesting because it allows perfect state transfer (a transfer of information with unit fidelity) independently of the chain length \cite{Christandl2004, nikolopoulos2004a} and is also exploited in other protocols, the latter is a reasonable trade-off between a good fidelity of transmission and the challenges to implement the model in a realistic experimental setup (the values considered here for $K$ are of the same order of $J$, so we are in a non-perturbative regime \cite{banchi2010}). An alternative method, exploiting local magnetic fields, has been proposed in \cite{lorenzo2013}. The interaction time $t$, in the implementation of this model in an integrated waveguide circuit, is proportional to the length of the circuit itself.

\section{\label{sec3}Qutrits and qubits: the average fidelity}

To begin with, let us analyze the approach in \cite{bose2003} and extend it to the case of qutrit transmission. As long as only a qubit has to be transferred and it is possible to initialize the chain with all the elements in their ground state, using a chain of bosons or fermions gives exactly the same results (by encoding, in the bosonic scenario, the two levels of the qubit as the ground and first excited state). The differences appear when more than one excitation is present in the whole chain; this could happen when the state to transmit requires more than two levels for its encoding (slightly different models have been studied in \cite{isart2007,bayat2007,ghosh2014,bayat2014}), as well as when there are unwanted excitations, due for instance to a non-perfect initialization of the medium. For the transfer of qutrits, we could encode the three levels as the ground and first two excited states, exploiting the possibility of having multiple excitations in the same channel. We suppose that the initial state of the system is separable and that all the qutrits but the first are in their ground state. Therefore, a general way to write the initial state reads
\begin{equation}
\ket{\Psi(0)}=(\alpha\ket{0}_1+\beta\ket{1}_1+\gamma\ket{2}_1)\ket{0}_{2}\cdots\ket{0}_N,
\end{equation}
where $\ket{1}_1$ indicates a single particle excitation in the first channel and $\ket{2}_1$ a double excitation corresponding to two particles in the first channel. From the definition of the Bloch sphere in five dimensions it follows \cite{caves2000}: $\alpha=\sin{\theta}\cos{\phi}e^{i\delta}$, $\beta=\sin{\theta}\sin{\phi}e^{i\sigma}$ and $\gamma=\cos{\theta}$, with the bounds $0\leq\theta\leq\frac{\pi}{2}$, $0\leq\phi\leq\frac{\pi}{2}$, $0\leq\delta\leq 2\pi$ and $0\leq\sigma\leq 2\pi$.

The evolved state at time $t$ will then be
\begin{equation}
\begin{split}
\ket{\Psi(t)}=&\alpha\ket{0}_{1}\cdots\ket{0}_N+\beta\sum_{j'=1}^N\left\langle j'|e^{-iHt}|1_{(1)}\right\rangle\ket{j'}+\\
+&\gamma\sum_{j''=1}^{N(N+1)/2}\left\langle j''|e^{-iHt}|2_{(1)}\right\rangle\ket{j''},
\end{split}
\end{equation}
where $\ket{1_{(1)}}=\ket{1}_1\ket{0}_{2}\cdots\ket{0}_N$ ($\ket{2_{(1)}}=\ket{2}_1\ket{0}_{2}\cdots\ket{0}_N$) corresponds to having one (two) photon(s) in the first channel and $\ket{j'}$ ($\ket{j''}$) is a general state in the subspace of single (double) excitation. By tracing out the states of channels from $1$ to $N-1$ it is then possible to get the density matrix $\rho_{out}^N$ of the qutrit in channel $N$ at time $t$ and thus the transfer fidelity. This, averaged over all possible initial states on a generalized Bloch sphere, reads
\begin{equation}
\begin{split}
\bar{F}&=\frac{2}{9\pi^2}\int_\Omega d\Omega \left\langle\Psi_{in}|\rho_{out}^N|\Psi_{in}\right\rangle\\
&=\frac{1}{3}+\frac{1}{12}|f_{N,1}|^2+\frac{1}{6}Re[f_{N,1}]+\frac{1}{12}|g_{\,2(N),2(1)}|^2\\
&+\frac{1}{6}Re[g_{\,2(N),2(1)}+f_{N,1}g^*_{\,2(N),2(1)}].\\
\end{split}
\end{equation}
Here, $f_{j,1}=\left\langle j| e^{-iHt}|1_{(1)}\right\rangle$ is the amplitude of finding a single photon in channel $j$ after time $t$ if a single excitation at time $t=0$ was in channel $1$. Similarly, $g_{\,i,2(1)}=\left\langle i| e^{-iHt}|2_{(1)}\right\rangle$ is the amplitude of having after time $t$ the double excitation corresponding to the state $i$ (in the double excitation basis) if two photons were in channel $1$ at time $t=0$. For the sake of completeness we observe that in order to compute $f$ and $g$ we need to apply, on the state the evolution is referred to, an excitation-dependent phase shift
\begin{equation}
\hat{\mathcal{R}}=
\begin{pmatrix}
1&0&0\\
0&e^{\frac{i\pi\cdot(N-1)}{2}}&0\\
0&0&e^{i\pi\cdot(N-1)}
\end{pmatrix}\;,
\end{equation}
$N$ being the number of channels. We remind the reader that, if we restrict our attention only to single particle excitations (qubits), the average fidelity is expressed by
\begin{equation}
\bar{F}_{1}=\frac{Re[f_{N,1}]}{3}+\frac{|f_{N,1}|^2}{6}+\frac{1}{2}.
\end{equation}

\begin{figure}[h]
\centering
\includegraphics[scale=0.29]{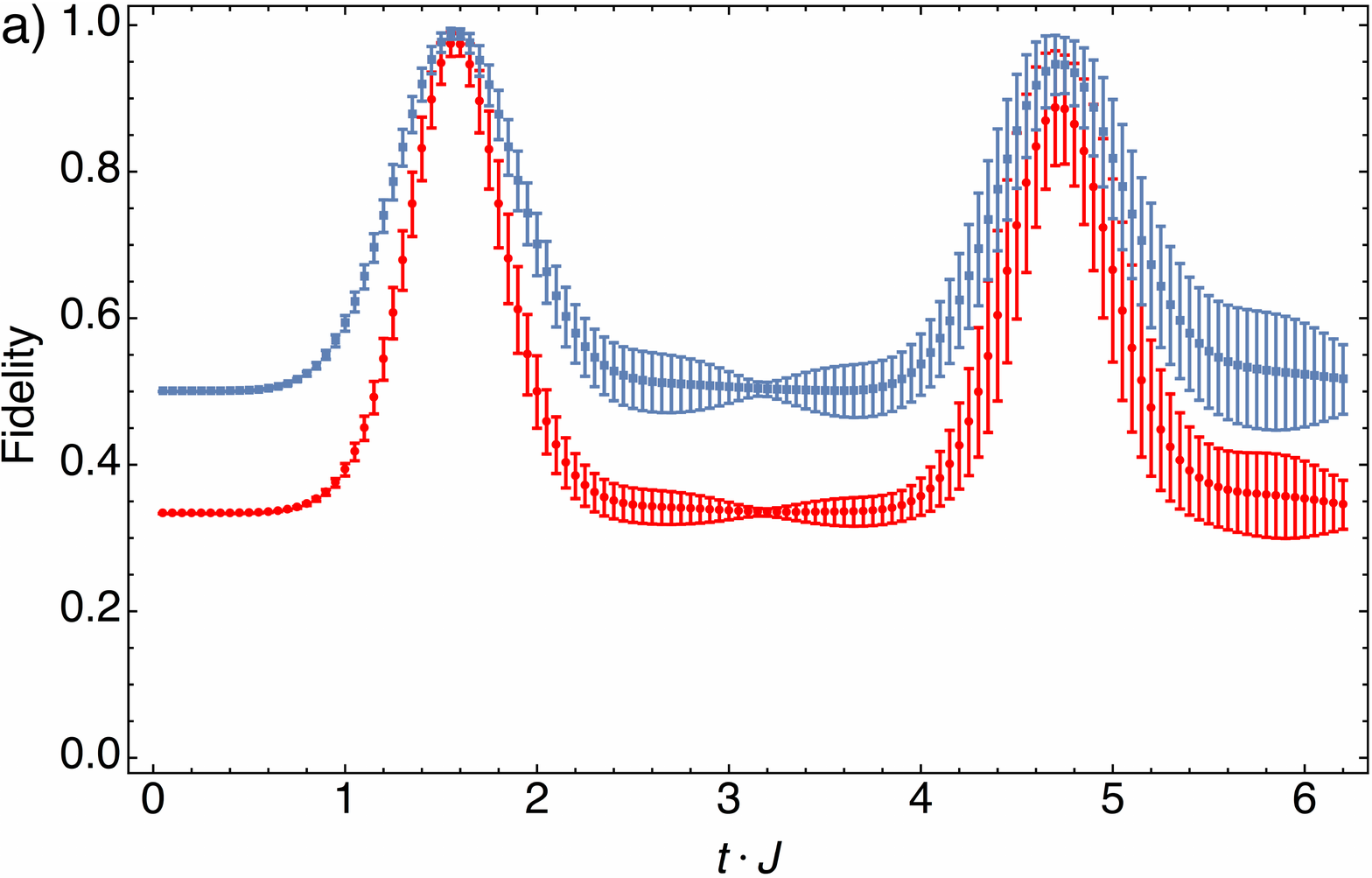}
\includegraphics[scale=0.29]{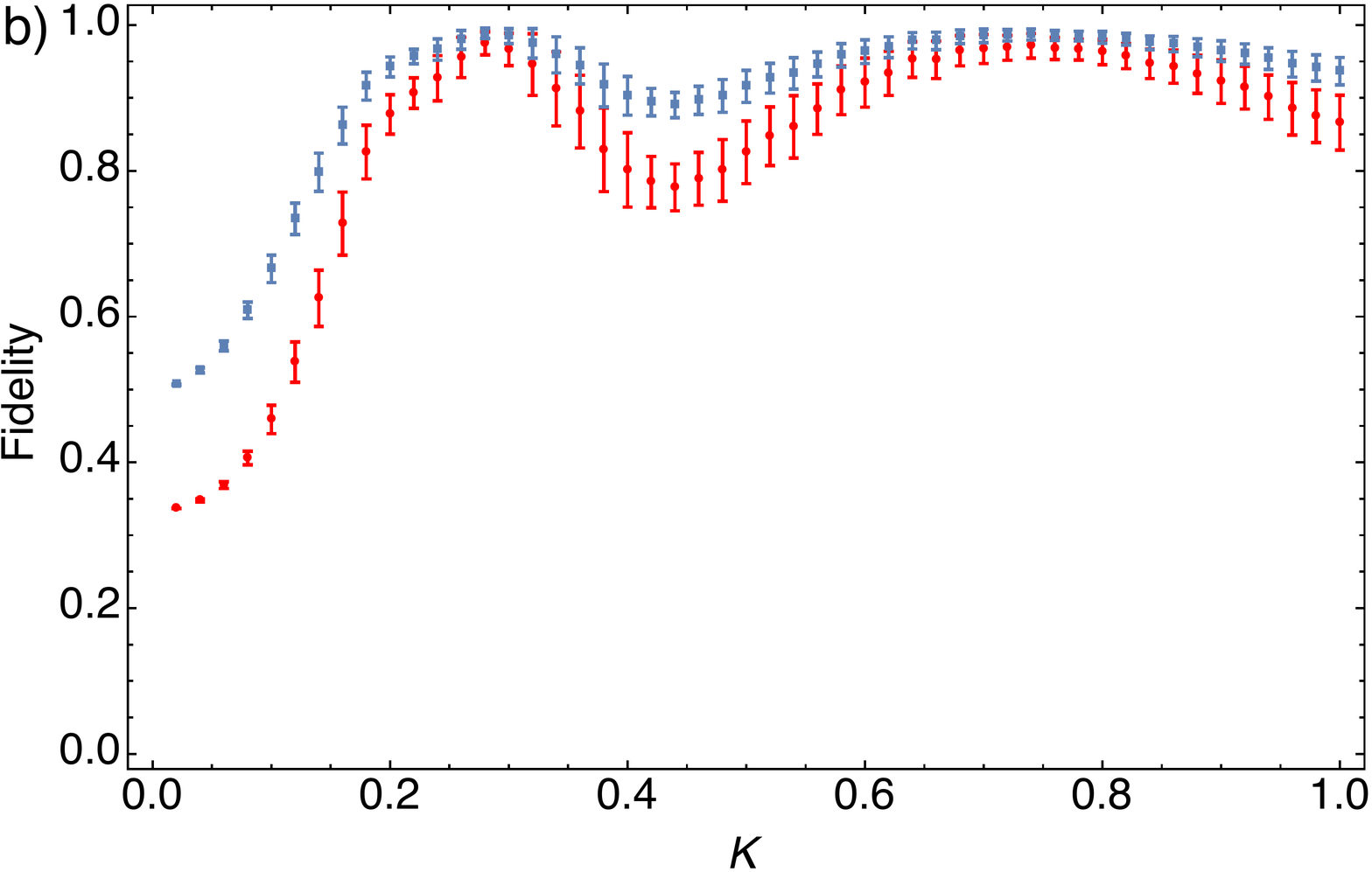}
\caption{Comparison between the fidelities for qubit (blue squares) and qutrit (red circles) transmission through a nine-channel chain. a) The case with mirror couplings $k_i=J\cdot\sqrt{i\cdot(N-i)}$; b) the case with equal couplings as a function of the edge couplings $K$ (maximizing over time in the range $t\cdot J\in[0,20]$).}
\label{uguali_bose}
\end{figure}

We display in Fig.\ref{uguali_bose} a comparison between the fidelities for qubit (blue squares) and qutrit (red circles) transmission through a nine-channel chain in the two scenarios described in Sec.~\ref{sec2}. The optimal values of $t$ and $K$ are the same for qubits and qutrits: due to the nature of the Hamiltonian in Eq.~(\ref{boson_hamiltonian}), the transfer of two excitations follows the same mechanisms of the transfer of a single excitation, as can also be understood from the information-flux analysis presented later on. Constant experimental errors on coupling strengths normally distributed and with variance $\sigma=5\%$ have been considered in the graphs. In other words, we have changed each $k_i$ according to $k_i\rightarrow k_i(1+\delta)$, with $\delta$ following a probability distribution $p(\delta)=(1/\sigma\sqrt{2\pi})e^{-\delta^2/2\sigma^2}$. This corresponds to a scenario where the distances between waveguides are not exactly the ideal ones. Moreover, by introducing an additional Gaussian error we analyzed the effects of time dependence in the couplings (thus taking into account also the case where the waveguides are not perfectly straight).

For a nine-channel system (with ideally equal couplings apart for the edge ones) we divided the evolution in $100$ steps each with a constant $4\%$ Gaussian error plus a further $2\%$ randomly distributed in time. Since fluctuations in time compensate, we obtained a peak in the fidelity transfer of $F_{qubit}=(0.987\pm 0.006)$ and $F_{qutrit}=(0.972\pm 0.012)$ respectively for qubits and qutrits, giving evidence of the system stability. It is interesting to notice that also the transfer of a general qutrit is very efficient in the regions where the fidelity for qubit transfer is close to $1$ (i.e. for the values of the parameters that one should use in an experiment to obtain a good transmission). One could expect that this holds for higher dimensions as well. Hence, for the transmission of information that would require more than a single qubit for its encoding, it is better to exploit the larger Hilbert space of the chain elements and encode it in qudits, instead of encoding in qubits and send them one by one (that would require a longer total time for the transfer) or using more elements of the chain \cite{lorenzo2015}.
In our analysis, we could have also considered the error due to photon losses. However, even if this is one of the main sources of error in some of the protocols exploiting waveguide circuits \cite{Sans2012, Crespi2012}, in principle we are assuming here only straight guides. Since we do not have bend losses, we expect the probability to lose a photon to be much lower.

To check the robustness of the protocol, we have simulated the transmission with a longer chain (up to $20$ channels), finding again very similar values for the transfer fidelity of qubits and qutrits. It is worth mentioning that the second scenario (equal couplings apart from the edge ones) shows more robustness compared with the case with mirror couplings, when we consider possible experimental errors (we have simulated also in this case errors up to $5\%$ for each coupling). This is in line with the results presented in \cite{Yung2005}, where it has been shown that more uniform couplings are more desirable.

\section{\label{sec4}Qubit transmission with extra excitations}
We would like now to investigate what happens to the transmission of qubits when the initialization of the medium is not perfect. In particular, we consider the scenario where each channel (apart from the first) has a probability $p$ of not being in its ground state. As we want to analyze this as a case of {\it noisy} transmission, we consider no coherence present among the different levels of each channel. We restrict our study to the scenario where only two possible levels of these channels can be initially populated. In the case of a spin chain this is always true, but for a bosonic chain this is an approximation that is still good for the values of probabilities that we are considering, if we think about a thermal state of each channel. All the elements of the chain will thus start in a statistical mixture $(1-p)\ket{0}\!\bra{0}+p\ket{1}\!\bra{1}$ apart from the first, that will be in the $\ket{\psi(0)}=\alpha\ket{0}_1+\beta\ket{1}_1$.

\begin{figure}[h]
\centering
\includegraphics[scale=0.4]{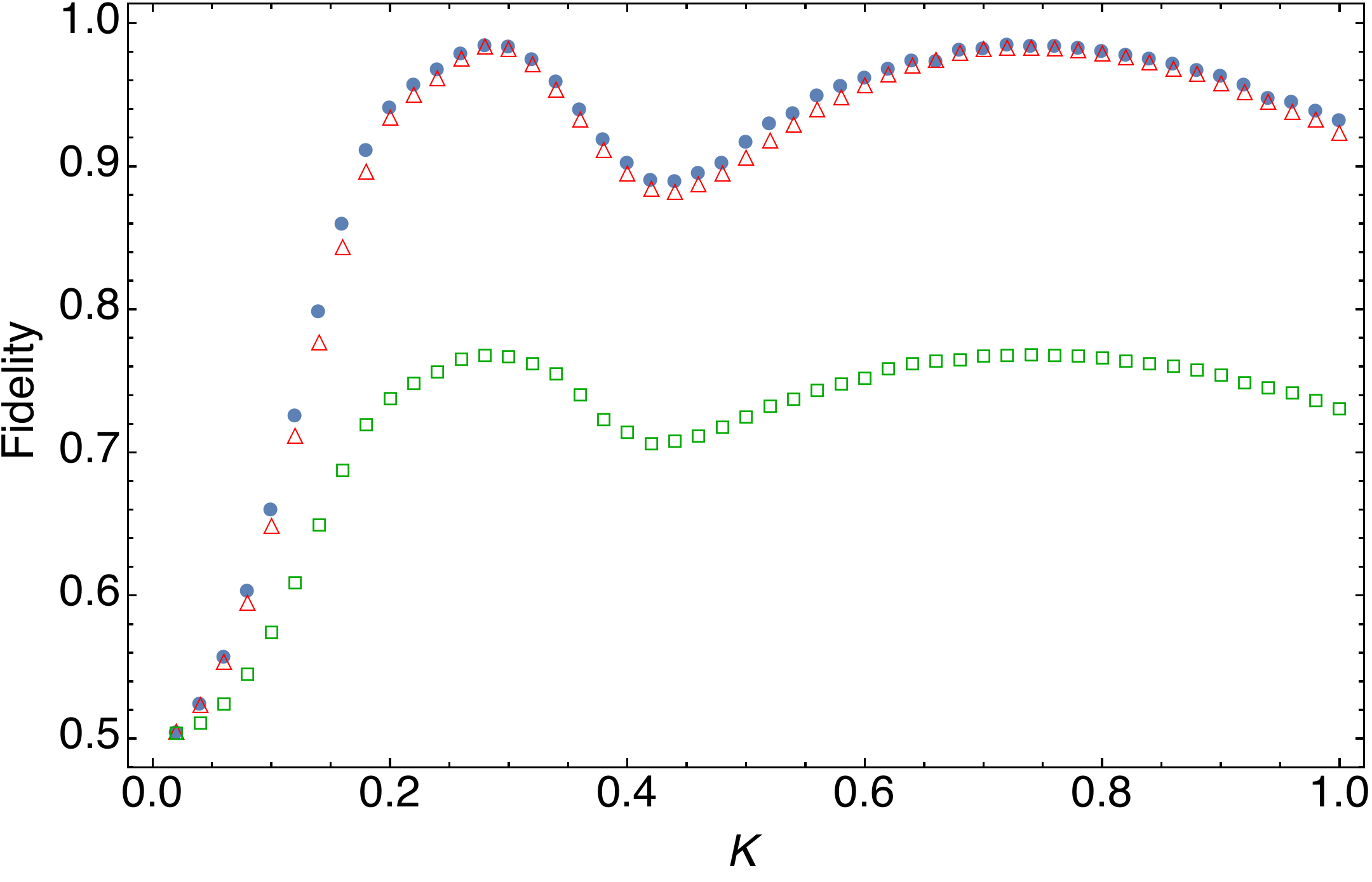}
\caption{Qubit fidelity transfer for a nine-channel chain. Experimental errors of $\sim 5\%$ on the couplings have been considered in all cases. Blue points: no extra-excitations; red triangles: a random extra-error excitation is averaged over all possible inputs in a bosonic chain; green squares: extra-error excitation in a  fermionic chain.}
\label{extraexcitation2}
\end{figure}

It is now interesting to distinguish between two main cases. The first is the transmission of a qubit in a chain that potentially could transmit qutrits (or even higher excitations), thus including the possibility of more excitations in a single channel (e.g. the evolution of a photon qubit plus extra photons in a waveguide circuit). On the other hand, for the qubit transmission through proper spin chains, we should discard the possibility of having multiple excitations in the same channel. By tracing out the states of channels from $1$ to $N-1$ and averaging over all possible initial states on the standard Bloch sphere, we can evaluate the average fidelity in both the cases. It is a key point to stress that in theory, if we knew a-priori whether input errors have occurred or not, we could apply different rotations on the output state: in the case of an extra excitation in a spin chain this corresponds to $\hat{\mathcal{R}}_{1-1}=\ket{0}\!\bra{0}+e^{i\pi N}\ket{1}\!\bra{1}$. The transmission fidelity would be very close to the ideal one both for fermionic and bosonic systems. Unfortunately, since we can not know the number of excitations on the chain, we will always apply the rotation corresponding to qubit transmission $\hat{\mathcal{R}}_{1}=\ket{0}\!\bra{0}+e^{i\pi (N-1)/2}\ket{1}\!\bra{1}$ which in general deeply affects the fidelity for both systems in the case of multiple excitations. However, in the very particular case of a bosonic chain with $4k+1$ channels ($k\in N$), or a fermionic chain with $4k+3$ channels ($k\in N$), the phase is revealed to be the same and thus there is no error transmission. As an example, in Fig.\ref{extraexcitation2} we compare boson and fermion fidelities for a nine channel chain in which each one of the remaining eight channels can have an extra input excitation with a probability $p_{extra}=0.05$. We can notice how the bosonic chain is much more convenient and less error sensitive for the transmission of information than its fermionic analogue in this case since the number of channels satisfies $N=4k+1$ ($k\in N$).

\section{\label{sec5}Information flux}

A quite different approach widely analyzed in \cite{Difranco2007,Difranco2008a,Difranco2008,Difranco2009} is called information flux: it consists in working in the Heisenberg picture, thus focusing the attention on the evolution of operators, rather than the specific input state. Indeed, we would like to understand how this information flux is related to the fidelity and in particular if it could be used to estimate the efficiency of the transmission in more general cases than those studied so far.

In the Heisenberg picture the evolution of an operator is regulated by $\mathcal{\hat{O}}_i(t)= \mathcal{\hat{U}^\dag} (t)\mathcal{\hat{O}}_i\mathcal{\hat{U}}(t)$, where $\mathcal{\hat{U}}(t) = \hat{T}exp[-(i/\hbar)\int\mathcal{\hat{H}}(t')dt']$. In the case of time independent Hamiltonian, the expression is simplified in $\mathcal{\hat{O}}_i(t)= e^{\frac{i}{\hbar}\mathcal{\hat{H}}t}\mathcal{\hat{O}}_ie^{-\frac{i}{\hbar}\mathcal{\hat{H}}t}$, which can be easily expanded as
\begin{equation}
\mathcal{\hat{O}}_i(t)=\mathcal{\hat{O}}_i + \frac{i}{\hbar}t[\mathcal{\hat{H},\hat{O}}_i] + \frac{1}{2}(\frac{i}{\hbar}t)^2[\mathcal{\hat{H}},[\mathcal{\hat{H},\hat{O}}_i]] + \cdots
\label{operator_evolution}
\end{equation}
As it has already been shown \cite{Difranco2007,Difranco2008a,Difranco2008,Difranco2009}, recalling \eqref{operator_evolution} and computing a series of commutators, it is possible to write the evolved operator $\mathcal{\hat{O}}_i(t)$ in terms of a set of time-dependent parameters $\mathcal{C} (t)$ and a set of operators $\mathcal{\hat{Q}}_j$ (all considered at time $t_0=0$), arising from the commutators.

Hence, we can now evaluate the time evolution of creation and annihilation operators in the case of bosons evolving with the Hamiltonian \eqref{boson_hamiltonian}. What we find out is that the set of commutators is composed respectively by all the creation and annihilation operators themselves. As a result, also the evolution can be expressed in the same terms:

\begin{equation}
\hat{a}^\dag_i(t)=\sum_{j=1}^N\mathcal{CR}_{(i,j)}(t)\hat{a}^\dag_j(0) \;,\; \hat{a}_i(t)=\sum_{j=1}^N\mathcal{AN}_{(i,j)}(t)\hat{a}_j(0).
\end{equation}

Given the evolution of a certain operator acting on the last channel of the chain, we define \emph{information flux} as the coefficient that sets the weight of the same operator applied at time zero on the first channel: $\mathcal{I}_N^{a(a^\dag)}=\mathcal{AN}(\mathcal{CR})_{(N,1)}$. We highlight that in this case $\mathcal{CR}(t)_{(i,j)}=\mathcal{AN}(t)_{(i,j)}\;\;\forall\; i,j,t$ for obvious symmetry reasons: for the sake of simplicity we will denote from now on as $\mathcal{C}_{(i,j)}$ both $\mathcal{CR}_{(i,j)}$ and $\mathcal{AN}_{(i,j)}$.

The evolution of the observable related to the number of photons is
\begin{equation}
\hat{a}^\dag_i(t)\hat{a}_i(t)=\sum_{j,k=1}^N\mathcal{CR}_{(i,j)}(t)\mathcal{AN}_{(i,k)}(t)\hat{a}^\dag_j \hat{a}_k
\end{equation}
and the corresponding information flux
\begin{equation}
\mathcal{I}_N^{a^\dag a}=\mathcal{CR}_{(N,1)}\mathcal{AN}_{(N,1)}.
\end{equation}

\begin{figure}[t!]
\centering
\includegraphics[scale=0.29]{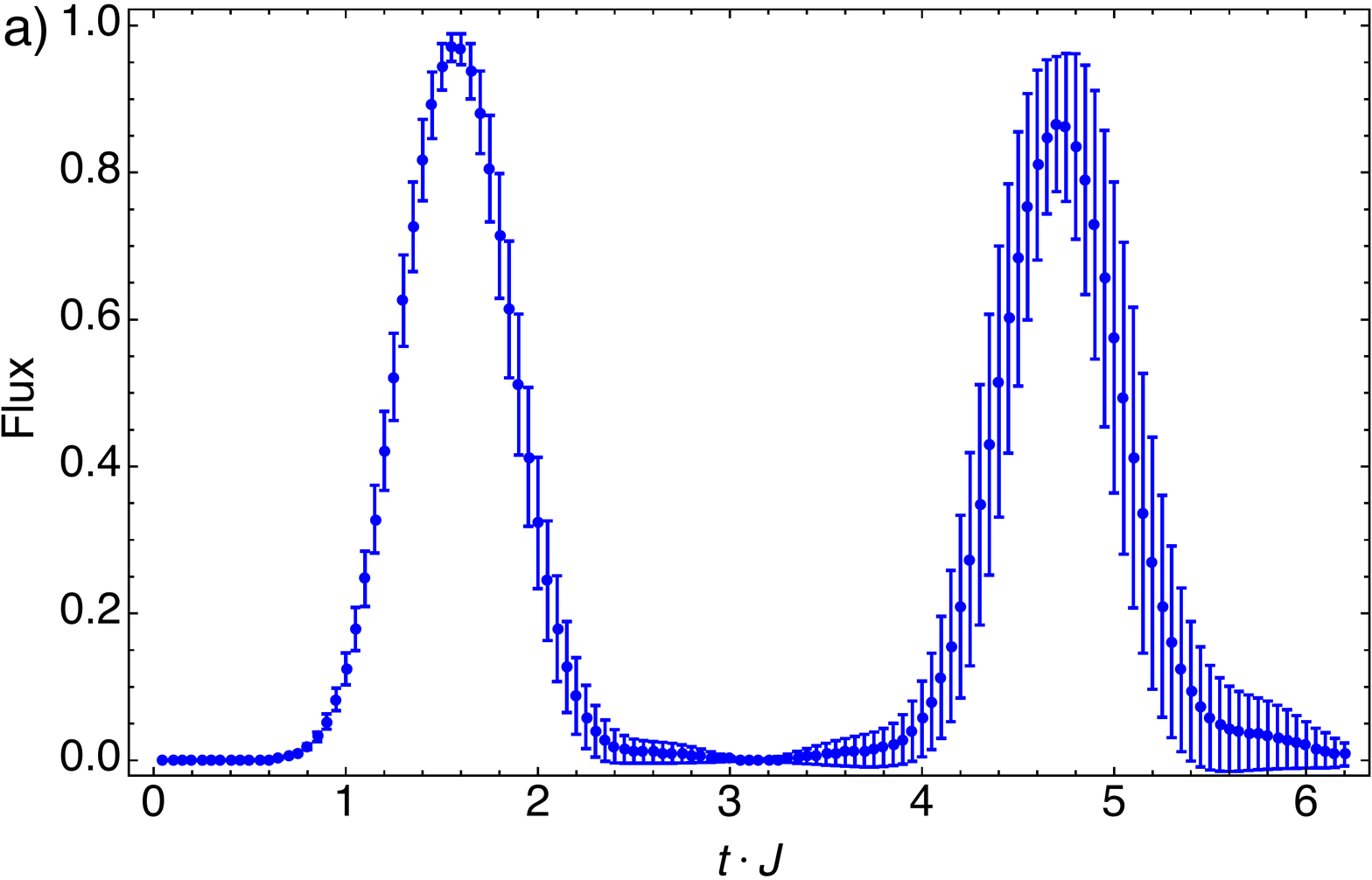}
\includegraphics[scale=0.29]{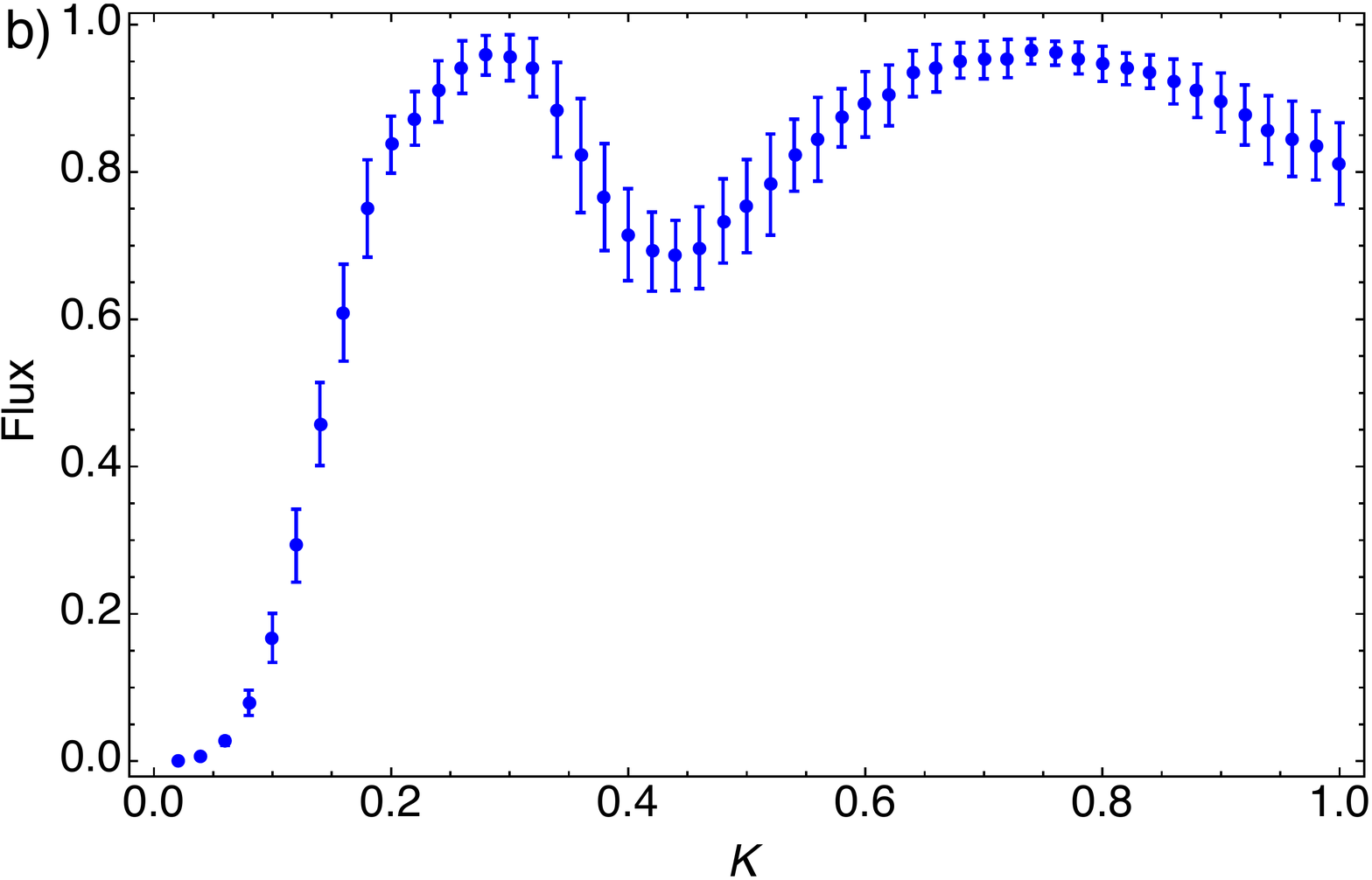}
\caption{Recalling Fig.\ref{uguali_bose} we plot the information flux $\mathcal{I}_N^{a^\dag a}$ for a nine-channel system in the case of mirror couplings (a) and for equal couplings (b) (maximizing over time in the range $t\cdot J\in[0,20]$). Experimental errors of $5\%$ on the couplings have been considered.}
\label{info_flux}
\end{figure}

In Fig.\ref{info_flux} we display the behavior of the information flux $\mathcal{I}_N^{a^\dag a}$ in a few relevant cases that we have already examined: the one with mirror couplings and with equal couplings (experimental errors of the order of $5\%$ on the couplings are taken into account).

\begin{figure}[h!]
\centering
\includegraphics[scale=0.3]{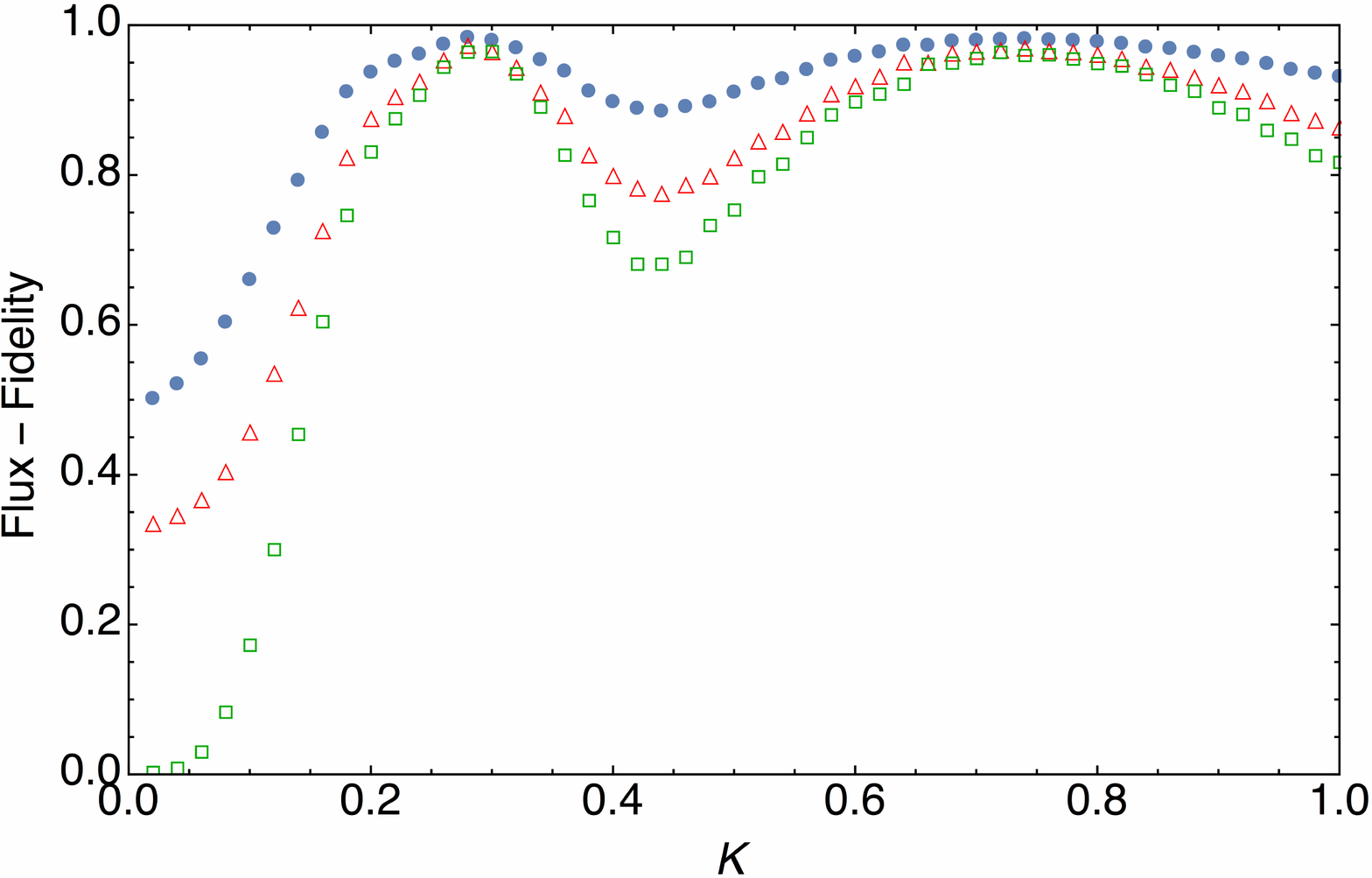}
\caption{Information flux $\mathcal{I}_N^{a^\dag a}$ (green squares) together with the transmission fidelity for qubits (blue circles) and qutrits (red triangles) in a linear chain with 9 channels (maximizing over time in the range $t\cdot J\in[0,20]$).}
\label{info_flux_fidelity}
\end{figure}

We remark that if we encode a qubit (qutrit) through the $\ket{0}$, $\ket{1}$ ($\ket{2}$) basis the number of photons in a channel is strictly related with the fidelity: we show in Fig.\ref{info_flux_fidelity} a comparison between information flux and fidelity for qubit and qutrits in a chain with equal couplings within a $5\%$ error. In addition, in the next section we would like to better understand how the $a^\dag_N a_N(t)$ operator that acts on the last channel at time $t$ is related with the information flux by considering various possible initial input states for channels $2\cdots N$. In particular, we will focus on Fock states, coherent states and a more general undefined state, though always considering the states in every channel pure and separable with respect to each other.

\subsection{Fock states}

Let us call $\ket{\Psi_{IN}(0)}$ the initial state of the system, being indeed a separable state where we have encoded our qubit in the first channel while every channel $i\geq 2$ is initially in the Fock state $\ket{i(0)}$. We would then have
\begin{equation}
\begin{split}
&n_n(t)=\left\langle\Psi_{IN}(0)|a^\dag_N a_N(t)|\Psi_{IN}(0)\right\rangle\\
&=\sum_{j,k}\mathcal{C}_{N,j}\mathcal{C}_{N,k}\left\langle\Psi_{IN}(0)|a^\dag_j a_k|\Psi_{IN}(0)\right\rangle\\
&=\sum_jC_{N,j}^2 n_j(0)\\
&=\mathcal{I}_N^{a^\dag a} n_1(0)+\sum_{j=2}^NC_{N,j}^2 n_j(0).
\end{split}
\end{equation}
The number of output photons in the last channel is thus given by the initial number of photons in the first channel, weighted by the information flux $\mathcal{I}_N^{a^\dag a}$, with a correction depending on the number of photons initially present in the other channels.

\subsection{Coherent states}

This time we suppose the system initially in a separable state where $\ket{\alpha_j}=e^{-|\alpha_j|^2/2}\sum_n\frac{\alpha_j^n}{\sqrt{n!}}\ket{n}_j$ is the initial coherent state in channel $j\geq 2$. We would then have
\begin{equation}
\begin{split}
&\left\langle\Psi_{IN}(0)|a^\dag_N a_N(t)|\Psi_{IN}(0)\right\rangle=\\
&\sum_{j,k}\mathcal{C}_{N,j}\mathcal{C}_{N,k}\left\langle\Psi_{IN}(0)|a^\dag_j a_k|\Psi_{IN}(0)\right\rangle = \sum_{j,k}\mathcal{C}_{N,j}\mathcal{C}_{N,k}\alpha_j^*\alpha_k\\
&=C_{N,1}^2 n_1(0) + \sum_{j=2}^NC_{N,j}^2 |\alpha_j|^2 + 2\sum_{j>k>1}Re[\alpha_j^*\alpha_k]C_{N,j}C_{N,k}+\\
&+2\sum_{j\neq 1}Re[\alpha_j^*]\sqrt{n_1(0)} C_{N,j}C_{N,1}.
\end{split}
\end{equation}
By manipulating these terms (as shown in the Appendix), we can find an upper bound for their weight, thus concluding that $\left\langle\Psi_{IN}(0)|a^\dag_N a_N(t)|\Psi_{IN}(0)\right\rangle=n_1(0)C^2_{N,1}+\Gamma$. Here $\Gamma$ represents the corrections which are bounded by
\begin{equation}
\begin{split}
\Gamma\leq &|\alpha_{max}|^2(1-C_{N,1}^2)(N-1)+\\
&+2\sqrt{n_1(0)} C_{N,1}|\alpha_{max}|\sqrt{N-1}\sqrt{(1-C_{N,1}^2)},
\end{split}
\end{equation}
where we point out that $C_{N,1}^2=\mathcal{I}_N^{a^\dag a}$.

\subsection{More general states}

Let us call $\ket{\Psi_{IN}(0)}$ the initial state (at time $t=0$):
\begin{equation}
\ket{\Psi_{IN}(0)}=\prod_{i=1}^N(\alpha_i(0)\ket{0}_i+\beta_i(0)\ket{1}_i+\gamma_i(0)\ket{2}_i+\cdots),
\end{equation}
where $i$ represents the channel and $\alpha (t)$, $\beta (t)$, $\gamma (t)$, $\cdots$ are the coefficients at time $t$ respectively of the vacuum component, one particle component, etc.

Now, recalling the general expression
\begin{equation}
\begin{split}
&\left\langle\Psi_{IN}(0)|a^\dag_N a_N(t)|\Psi_{IN}(0)\right\rangle=\\
&=\sum_{j,k}\mathcal{C}_{N,j}\mathcal{C}_{N,k}\left\langle\Psi_{IN}(0)|a^\dag_j a_k|\Psi_{IN}(0)\right\rangle,
\end{split}
\end{equation}
it is possible to split it into two parts
\begin{equation}
\left\langle\Psi_{IN}(0)|a^\dag_j a_j|\Psi_{IN}(0)\right\rangle = n_j(0)
\end{equation}
and
\begin{equation}
\begin{split}
&\left\langle\Psi_{IN}(0)|a^\dag_j a_k|\Psi_{IN}(0)\right\rangle |_{j\neq k}=\\
&=(\beta_k(0)\alpha_k^*(0)+\sqrt{2}\gamma_k(0)\beta_k^*(0)+\sqrt{3}\delta_k(0)\gamma_k^*(0)+\cdots)\cdot\\
&\cdot(\alpha_j(0)\beta_j^*(0)+\sqrt{2}\beta_j(0)\gamma_j^*(0)+\sqrt{3}\gamma_j(0)\delta_j^*(0)+\cdots).
\end{split}
\end{equation}
While the first term is easy to calculate, being actually very similar to the case of coherent states, instead, in order to maximize the second term, we observe that each one of the two parentheses is the result of the product $_{i}\left\langle i(0)|a|i(0)\right\rangle_{i}$ on the initial state of channel $i$. Since the state $\ket{i(0)}$ is normalized in the Fock basis, this quantity is upper-bounded by $1$, and it is actually equal to one only in the case of a coherent state, which therefore reveals to be the worst case. This means that the second term can be bounded by:
\begin{equation}
\left\langle\Psi_{IN}(0)|a^\dag_j a_k|\Psi_{IN}(0)\right\rangle |_{j\neq k}\leq \sum_{j\neq k} C_{N,j}C_{N,k}\sqrt{n_j}\sqrt{n_k}
\end{equation}
Following exactly the same procedure of the previous paragraph, we can finally conclude that in the general case $\left\langle\Psi_{IN}(0)|a^\dag_N a_N(t)|\Psi_{IN}(0)\right\rangle=C^2_{N,1}\cdot n_1(0)+\Gamma$. Here, $\Gamma$ represents the corrections which are bounded by
\begin{equation}
\begin{split}
&\Gamma\leq n_j(0)_{max}(1-C_{N,1}^2)(N-1)+\\
&2\sqrt{n_1(0)} C_{N,1}\sqrt{n_j(0)_{max}}\sqrt{N-1}\sqrt{(1-C_{N,1}^2)},
\end{split}
\end{equation}
where again $C_{N,1}^2=\mathcal{I}_N^{a^\dag a}$.\\

\section{\label{sec6}Conclusions}
The rapid and striking progress obtained in the field of integrated circuits has allowed the experimental implementation of several quantum information protocols. Another promising direction for exploiting these setups would be the adaptation of schemes that have been already proposed in literature for spin chains. Considering their wide range of applications, from the faithful transmission of information to its manipulation and control, including the possibility of generating quantum entanglement, this will clearly pave the way to further progress. However, we have to keep in mind that, even if the map from a fermionic to a bosonic chain is exact in some particular scenario, this is not always the case. One can also find advantages in using these adapted protocols with respect to the standard spin-chain schemes. For instance, here we have investigated the possibility of transmitting three-level quantum states by encoding them in the multiple channel-excitations of a bosonic chain, with a fidelity very close to the case of qubits in spin chains. Moreover, we have analyzed the effect on the transfer fidelity of unwanted excitations, that could happen in a realistic experimental implementation. We have highlighted how the cases of bosonic and fermionic chains give very similar results, but they could strongly differ if the number of channels satisfies particular conditions. Finally, an analysis from a different viewpoint, namely an approach based on the information flux, has allowed us to obtain bounds to the fidelity in more general scenarios.

\section{\label{sec7}Acknowledgments}
We thank the UK EPSRC (EP/K034480/1) for support.

\appendix
\begin{widetext}
\section{Coherent states}
Hereafter we report the calculations of the correction term for information flux in the case of coherent states in input.
\begin{itemize}
\item
\begin{equation}
\sum_{j=2}^NC_{N,j}^2 |\alpha_j|^2 \leq |\alpha_j^{max}|^2\cdot(1-C_{N,1}^2)
\end{equation}

\item
\begin{equation}
\begin{split}
&2\sum_{j>k>1}Re[\alpha_j^*\alpha_k]C_{N,j}C_{N,k} \leq |\alpha_{max}|^2\sum_{j\neq k\neq 1}|C_{N,j}||C_{N,k}|\\
&\leq |\alpha_{max}|^2\cdot\sqrt{(1-C_{N,1}^2)}\sqrt{N-1}\cdot\frac{\sqrt{(1-C_{N,1}^2)}}{\sqrt{N-1}}\cdot (N-2)\\
&=|\alpha_{max}|^2(1-C_{N,1}^2)(N-2)
\end{split}
\end{equation}
where, since $\sum_j|C_{N,j}|^2=1$, the maximum of the sum $\sum_{j\neq 1}|C_{N,j}|$ is obtained in the case in which the coefficients are all the same and equal to $|C_{N,j}|=\sqrt{\frac{(1-C_{N,1}^2)}{N-1}}$. In addition, the second inequality holds since the second sum covers $N-2$ terms, while the first sum is over $N-1$ addends.
\item
\begin{equation}
\begin{split}
&2\sqrt{n_1(0)}\sum_{j\neq 1}Re[\alpha_j^*]C_{N,j}C_{N,1}=2\sqrt{n_1(0)}C_{N,1}\cdot\sum_{j\neq 1}\alpha_j^*C_{N,j}\\
&\leq 2\sqrt{n_1(0)} C_{N,1}|\alpha_{max}|\sum_{j\neq 1} |C_{N,j}|\\
&\leq 2\sqrt{n_1(0)} C_{N,1}|\alpha_{max}|\sqrt{N-1}\sqrt{(1-C_{N,1}^2)}
\end{split}
\end{equation}
where again, since $\sum_j|C_{N,j}|^2=1$, the maximum of the sum $\sum_{j\neq 1}|C_{N,j}|$ is obtained in the case in which the coefficients are all the same and equal to $|C_{N,j}|=\sqrt{\frac{(1-C_{N,1}^2)}{N-1}}$.
\end{itemize}

\end{widetext}

\bibliographystyle{apsrev4-1}
\bibliography{Draft_fidelity_flux}

\end{document}